\documentclass[pra,twocolumn,showpacs,superscriptaddress,floatfix, reprint]{revtex4}

\usepackage[OT1]{fontenc}
\usepackage[usenames,dvipsnames]{color}
\usepackage[latin1]{inputenc}
\usepackage[english]{babel}
\usepackage{graphicx}
\usepackage{color}
\usepackage{epstopdf}
\usepackage[Gray,squaren]{SIunits}
\usepackage{xspace}
\usepackage{upgreek}
\usepackage{ulem}
\usepackage{epstopdf}
\usepackage{amssymb,amsmath,verbatim,ulem}

\newcommand{\new}[1]{\textcolor{blue}{#1}}

\newcommand{\ket}[1]{|{#1}\rangle}
\newcommand{\bra}[1]{\langle{#1}|}

\begin{document}

\title{Micro-wave assisted transparency in a M-system }
\author{Nawaz Sarif Mallick}
\author{ Tarak N. Dey}
\author{Kanhaiya Pandey}
\affiliation{Department of Physics, Indian Institute of Technology Guwahati, Guwahati, Assam 781039, India }
 \email{kanhaiyapandey@iitg.ernet.in}

\date{\today{}}

\begin{abstract}
   In this work we theoretically study a five level M-system whose two unpopulated ground states are coupled by a micro-wave (MW) field. 
  The key feature which makes M-systems more efficient in comparison to  closed loop $\Lambda$ systems is the absence of MW field induced population transfer even at high intensities of the later.
   We examine lineshape of probe absorption as a function of its detuning in the presence of both control and MW fields. 
   The MW field facilitates the narrowing of the probe absorption lineshape in M-systems which is in contrast to closed loop $\Lambda$-systems.
   Hence this study opens up a new avenue for atom based phase dependent MW magnetometry.
\end{abstract}


\maketitle

\section{Introduction}

Recently there are efforts towards atom based microwave (MW) electro- and magneto-metry due to their high
reproducibility, accuracy, resolution and stability \cite{SSK13, JSA12, JYR16}. This is based upon the phenomenon of electromagnetically induced transparency (EIT)
\cite{EIT, Bo1} in which the absorption property of a probe laser is altered in the presence of control lasers in a multilevel system. 
The basis of EIT is the induced or transfer of coherence, TOC between the levels which are not directly driven by optical control fields. 
The EIT has been explored extensively in the three level $\Lambda$ \cite{MSL11, IKN08,IFN09,Kimball,Min1,Phillips,TNDey},
V \cite{MEA99, V2,V4} and $\Xi$ \cite{KPW05,JDP,AVG,Schmidt,L1,JIN2,Sumanta}
 systems. It has also been investigated beyond three level systems such as in doubly driven V \cite{EGD01}, N \cite{Slow, BMW08,GWR04,CXH09,CSB11}, Y \cite{TLY15}, inverted Y \cite{LBF12}, $\Xi\Lambda$ 
\cite{HCW05,PKN11}, tripod \cite{KLG11, SLB13} and doubled tripod \cite{HMW14} systems. The role of various TOCs for a general $N$-level atomic system has been well
studied \cite{PAN13}. This phenomenon has been paid a great deal of attention due to its potential applications in a wide variety of fields like controlling
the group velocity of light \cite{Kimball,Slow}, coherent storage and retrieval of light \cite{Phillips,TNDey},
high-resolution spectroscopy \cite{KPW05, JIN2},
studying photon-photon interaction via Rydberg blockade \cite{JDP, AVG}, photon transistor for quantum optical information processing \cite{Schmidt, Khazali,Sumanta}, etc.

The induced coherence of a $\Lambda$ system can be drastically modified by applying a low frequency field which couples  two ground levels. Two optical fields along with a
lower level coupling field (LL) form a closed loop three level $\Lambda$ system. The relative amplitude, detuning  and phase differences between the LL field and two
optical fields can play an important role in significantly manipulating the optical property of the system such as dispersion, absorption and nonlinearity
\cite{TND,LVY09,MUA14,RTS15,Eilam:09,KMR92,RCP16}.

In the above mentioned closed loop $\Lambda$ system, population transfer is unavoidable in the presence of high power MW
field as it acts on one of the populated states. 
The MW induced population transfer between metastable states 
leads to imperfect transparency which  limits many of the EIT based applications \cite{LUK03}.
Here we study an M-system wherein two unoccupied ground states are coupled with a MW field and hence there is no population transfer with this MW Rabi frequency.
This MW field also enhances the generated atomic coherence at the probe transition manifold which can not be found in generic three level systems.
This enhancement of atomic coherence is a result of the interference between excitation pathways from the different atomic states.

The paper is organized as follows : 
in the next section, we introduce the physical model and basic equations of motion for a M-system by using a semiclassical theory.
In section \ref{Perturbative Analysis}, an approximate analytical expression for linear susceptibility of a weak probe field is derived using the perturbative approach under three photon resonance condition.
After deriving the analytical form of the atomic coherences, we compare it with the full numerical solution.
In section \ref{RESULTS}, we first explain the lineshape of the probe laser absorption as a function of probe detuning for various MW strengths and phases.
Finally M-systems are compared to closed loop $\Lambda$-systems both having ground states coupled by MW field.

\section {Theoretical formulation}
\label{Theoretical formulation}
\subsection {Model Configuration}
\label{Model Configuration}

\begin{figure}
   \begin{center}
      \includegraphics[width =\linewidth]{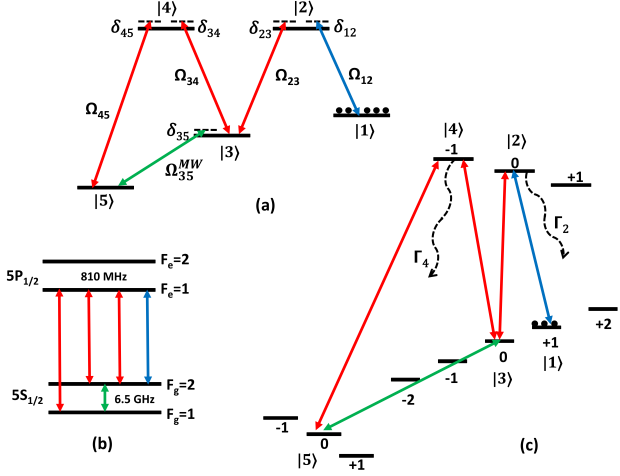}
      \caption{(Color online). (a) The general energy level diagram for M-system. (b) The relevent energy level of $^{87}$Rb for D$_1$ line to realize the M-system. (c) The
magnetic sublevels in the presence of the magnetic field of the hyperfine states to realize the M-system.}
      \label{Msysenelev}
   \end{center}
\end{figure}

We consider a five level atomic M-system as shown in Fig. \ref{Msysenelev}a. The electric field associated with the lasers driving the transition $\ket{i}$ $\rightarrow$ $\ket{i+1}$
is E$_{i, i+1}e^{i(\omega^L_{i, i+1}t+\phi^L_{i, i+1})}$. The E$_{i, i+1}$ is amplitude, $\omega_{i, i+1}$ is the frequency and $\phi^L_{i, i+1}$ is the phase. We define
Rabi frequency $\Omega_{i,j} =d_{i,j}E_{i,j}e^{i\phi^L_{i,j}}/\hbar$ for the transition $\ket{i}$ $\rightarrow$ $\ket{j}$ having the dipole moment matrix element $d_{i,j}$
driven by a laser with electric field amplitude E$_{i,j}$ and the phase $\phi^L_{i,j}$ . The $\Omega_{12}$ is the Rabi frequency of the probe laser driving $\ket{1}$
$\rightarrow$ $\ket{2}$ and shown with solid blue arrow in Fig. \ref{Msysenelev}. 
The  Rabi frequencies of the control lasers driving the transitions $\ket{2}$ $\rightarrow$
$\ket{3}$, $\ket{3}$ $\rightarrow$ $\ket{4}$ and $\ket{4}$ $\rightarrow$ $\ket{5}$ are denoted by $\Omega_{23}$, $\Omega_{34}$ and $\Omega_{45}$ respectively and shown with solid red arrows. 
The lower level states $\ket{3}$ and  $\ket{5}$ is coupled by a MW field with Rabi frequency  $\Omega^{MW}_{35}$ as shown with solid green arrow.
The M-system can be realized in $^{87}$Rb at the D$_1$ line using the two hyperfine levels $F_g=1$ and $F_g=2$ of the ground state 5S$_{1/2}$ and the hyperfine level
$F_e=1$ of excited state 5P$_{1/2}$ as shown in Fig. \ref{Msysenelev}b. The decay rate of the excited state 5P$_{1/2}$ is $2\pi\times$6~MHz. We apply a large magnetic
field to split the magnetic sublevels to keep away the unwanted sublevels from resonance and to define the quantization axis. The sign and value of the Lande g factor for
the chosen hyperfine is suitable to split and keep away the unwanted transitions. At 200~G applied magnetic field the splitting of the magnetic sublevels for $F_e=1$ is 2$\pi\times$46
MHz and splitting for $F_g=2$ and $F_g=1$ is 2$\pi\times$140 MHz as shown in Fig. \ref{Msysenelev}c.  We consider the probe to be $\sigma^-$ polarized and driving the transition  $\ket{F_g=2;m_F=+1}$ $\rightarrow$
$\ket{F_e=1;m_F=0}$. The $\pi$ polarized control laser is driving the $\ket{F_g=2;m_F=0}$ $\rightarrow$ $\ket{F_e=1;m_F=0}$ transition. The two $\sigma^-$ polarized control
lasers are driving the  $\ket{F_g=2;m_F=0}$ $\rightarrow$ $\ket{F_e=1;m_F=-1}$ and $\ket{F_g=1;m_F=0}$ $\rightarrow$ $\ket{F_e=1;m_F=-1}$ transitions as shown in Fig. 
\ref{Msysenelev} c.
   
In the absence of the control lasers all the magnetic sublevels of the hyperfine states $F_g=1$ and $F_g=2$ will be equally populated. However when the control lasers
are applied\new{,} the population in the ground states connected with the control lasers will deplete. In the steady state there will be no population left
in the $\ket{F_g=1,m_F=0}$ and $\ket{F_g=2,m_F=0}$ states as shown in Fig. \ref{Msysenelev}c.               

The unperturbed atomic Hamiltonian can be written as
\begin{equation}
H_0=\sum_{j=1}^{5}\hbar\omega_j\ket{j}\bra{j},
\end{equation}
where $\hbar\omega_{j}$ is the energy of the $\ket{j}$ state. Under the action of five coherent fields, the interaction Hamiltonian of the system in the dipole approximation is given by
\begin{align}
&H_I=\frac{\hbar\Omega_{12}}{2}[e^{i\omega^L_{12}t}+e^{-i\omega^L_{12}t}]\ket{1}\bra{2}\nonumber \\
&+\frac{\hbar\Omega_{23}}{2}[e^{i\omega^L_{23}t}+e^{-i\omega^L_{23}t}]\ket{2}\bra{3} \nonumber \\
&+\frac{\hbar\Omega_{34}}{2}[e^{i\omega^L_{34}t}+e^{-i\omega^L_{34}t}]\ket{3}\bra{4}\nonumber \\
&+\frac{\hbar\Omega_{45}}{2}[e^{i\omega^L_{45}t}+e^{-i\omega^L_{45}t}]\ket{4}\bra{5} \nonumber\\
&+\frac{\hbar\Omega^{MW}_{35}}{2}[e^{i\omega^{MW}_{35}t}+e^{-i\omega^{MW}_{35}t}]\ket{3}\bra{5}
\end{align}
Hence the total Hamiltonian will be $H=H_0+H_I$.
We use suitable unitary transformation to eliminate the explicit time dependent part in the Hamiltonian. 
In rotating wave approximation, the effective interaction Hamiltonian can be expressed as 
\begin{align}
&{\mathcal H_I}=\hbar[0\ket{1}\bra{1}-\delta_{12}\ket{2}\bra{2}-(\delta_{12}-\delta_{23})\ket{3}\bra{3}\nonumber \\
&-(\delta_{12}-\delta_{23}+\delta_{34})\ket{4}\bra{4}-(\delta_{12}-\delta_{23}+\delta_{34}-\delta_{45})\ket{5}\bra{5}\nonumber \\
&+\frac{\Omega_{12}}{2}\ket{1}\bra{2}+\frac{\Omega_{23}}{2}\ket{2}\bra{3}+\frac{\Omega_{34}}{2}\ket{3}\bra{4}+\frac{\Omega_{45}}{2}\ket{4}\bra{5}\nonumber\\
&+\frac{\Omega^{MW}_{35}}{2}e^{i(-\delta_{34}+\delta_{45}+\delta^{MW}_{35})t}\ket{3}\bra{5}]+c.c.,
\end{align}
where $\delta_{12}=\omega^L_{12}-(\omega_2-\omega_1)$, $\delta_{23}=\omega^L_{23}-(\omega_2-\omega_3)$, $\delta_{34}=\omega^L_{34}-(\omega_4-\omega_3)$,
$\delta_{45}=\omega^L_{45}-(\omega_4-\omega_5)$ and $\delta^{MW}_{35}=\omega^{MW}_{35}-(\omega_5-\omega_3)$ known as detunings of the lasers for respective transitions.
Now we impose the condition $\delta_{45}+\delta_{35}^{MW}-\delta_{34}=0$, so that the time dependence is completely eliminated from the effective interaction Hamiltonian.
To explore the dynamics of the atomic system, we use the density matrix equations where radiative relaxation of the atomic states are included. 
The atomic density operator $\rho$ obey the following equations
\begin{equation}
\dot{\rho}=-\frac{i}{\hbar}[{\mathcal H_I}, \rho]-\frac{1}{2}\{\Gamma,\rho\}
\label{ME}
\end{equation}
where $\Gamma$ is the relaxation matrix \cite{scully1997}.

\subsection {Perturbative Analysis}
\label{Perturbative Analysis}

\begin{figure}
   \begin{center}
      \includegraphics[width =\linewidth]{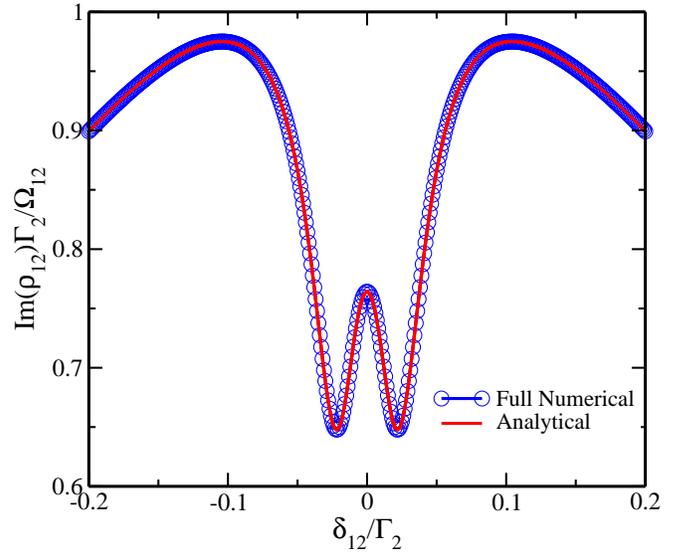}
      \caption{(Color online). Comparison between analytical and numerical solution for normalized absorption, $\textrm{Im}(\rho_{12})\Gamma_2/\Omega_{12}$ vs $\delta_{12}/\Gamma_2$. The red solid curve corresponds to the analytical solution while blue open circle points correspond to the full numerical solution for $\Omega_{12}$=$2\pi\times$ 0.01~MHz, $|\Omega_{23}|$=$|\Omega_{34}|$=$|\Omega_{45}|$=$2\pi\times$1~MHz, $|\Omega^{MW}_{35}|$=$2\pi\times$0.3~MHz, $\phi$=$\pi$/2, $\delta_{23}$=$\delta_{34}$=$\delta_{45}$=$\delta^{MW}_{35}$=0, $\Gamma_1=\Gamma_3=\Gamma_5=0$  and $\Gamma_2=\Gamma_4=2\pi\times$ 6~MHz.}
      \label{numvsann}
   \end{center}
\end{figure}

To study the response of the medium, we numerically solve the density matrix equations at steady-state condition.
We derive an analytical expression for the probe field response that is correct to first order in  probe field
approximation and exact for the all order in  control fields.
In this weak approximation, there will be no population transfer and hence the evolution of the population i.e. diagonal terms
of the density matrix such as
$\rho_{11}$, $\rho_{22}$, $\rho_{33}$, $\rho_{44}$ and $\rho_{55}$ can be ignored with $\rho_{11}\approx1$,  $\rho_{22}\approx0$, $\rho_{33}\approx0$, $\rho_{44}\approx0$
and $\rho_{55}\approx0$. Time evolution for the coherences i.e. off-diagonal terms of the density matrix will be given by following set of equations
\begin{align}
\dot{\rho}_{12}&\approx-\gamma_{12}\rho_{12}+\frac{i}{2}\Omega_{12}+\frac{i}{2}\Omega^*_{23}\rho_{13},\nonumber\\
\dot{\rho}_{13}&\approx-\gamma_{13}\rho_{13}-\frac{i}{2}\Omega_{12}\rho_{23}+\frac{i}{2}\Omega_{23}\rho_{12}\nonumber\\
&+\frac{i}{2}\Omega^*_{34}\rho_{14}+\frac{i}{2}\Omega^{MW*}_{35}\rho_{15},\nonumber\\
\dot{\rho}_{14}&\approx-\gamma_{14}\rho_{14}-\frac{i}{2}\Omega_{12}\rho_{24}+\frac{i}{2}\Omega_{34}\rho_{13}+\frac{i}{2}\Omega^*_{45}\rho_{15},\nonumber\\
\dot{\rho}_{15}&\approx-\gamma_{15}\rho_{15}-\frac{i}{2}\Omega_{12}\rho_{25}+\frac{i}{2}\Omega^{MW}_{35}\rho_{13}+\frac{i}{2}\Omega_{45}\rho_{14},
\label{coherence}
\end{align}

where 
\begin{align}
\gamma_{12}&=\left[\frac{\Gamma_{1}+\Gamma_{2}}{2}+i\delta_{12}\right],\nonumber\\
\gamma_{13}&=\left[\frac{\Gamma_{1}+\Gamma_{3}}{2}+i\left(\delta_{12}-\delta_{23}\right)\right],\nonumber\\
\gamma_{14}&=\left[\frac{\Gamma_{1}+\Gamma_{4}}{2}+i\left(\delta_{12}-\delta_{23}+\delta_{34}\right)\right],\nonumber\\
\gamma_{15}&=\left[\frac{\Gamma_{1}+\Gamma_{5}}{2}+i\left(\delta_{12}-\delta_{23}+\delta_{34}-\delta_{45}\right)\right].\nonumber
\end{align}
The decay rate of the states $\ket{1}$, $\ket{2}$, $\ket{3}$, $\ket{4}$ and $\ket{5}$ are given by $\Gamma_{1}$, $\Gamma_{2}$, $\Gamma_{3}$, $\Gamma_{4}$ 
and $\Gamma_{5}$, respectively.  
Again within the same weak probe limit, the coherences $\rho_{23}$ $\approx$$\rho_{24}$$\approx$$\rho_{25}$$\approx$0. 
Under the steady state condition i.e. $\dot{\rho}_{ij}=0$, the equations (\ref{coherence}) become

\begin{align}
\rho_{12}&=\frac{i}{2}\frac{\Omega_{12}}{\gamma_{12}}+\frac{i}{2}\frac{\Omega^*_{23}}{\gamma_{12}}\rho_{13}\nonumber\\
\rho_{13}&=\frac{i}{2}\frac{\Omega_{23}}{\gamma_{13}}\rho_{12}+\frac{i}{2}\frac{\Omega^*_{34}}{\gamma_{13}}\rho_{14}+\frac{i}{2}\frac{\Omega^{MW*}_{35}}{\gamma_{13}}\rho_{15}\nonumber\\
\rho_{14}&=\frac{i}{2}\frac{\Omega_{34}}{\gamma_{14}}\rho_{13}+\frac{i}{2}\frac{\Omega^*_{45}}{\gamma_{14}}\rho_{15}\nonumber\\
\rho_{15}&=\frac{i}{2}\frac{\Omega^{MW}_{35}}{\gamma_{15}}\rho_{13}+\frac{i}{2}\frac{\Omega_{45}}{\gamma_{15}}\rho_{14}
\label{cohsteady}
\end{align}
We obtain following analytical expression for $\rho_{12}$ by solving above linear algebraic equations
\begin{align}
\rho_{12}&=\frac{\frac{i}{2}\frac{\Omega_{12}}{\gamma_{12}}}{1+\frac{\frac{1}{4}\frac{|\Omega_{23}|^2}{\gamma_{12}\gamma_{13}}}{1+\frac{\frac{1}{4}\frac{|\Omega_{34}|^2}{\gamma_{13}\gamma_{14}}+\frac{i}{4}\frac{|\Omega_{34}||\Omega_{45}||\Omega^{MW}_{35}|cos\phi}{\gamma_{13}\gamma_{14}\gamma_{15}}+\frac{1}{4}\frac{|\Omega^{MW}_{35}|^2}{\gamma_{13}\gamma_{15}}}{1+\frac{1}{4}\frac{|\Omega_{45}|^2}{\gamma_{14}\gamma_{15}}}}},
\label{ana}
\end{align}
where $\phi=\phi^{MW}_{35}-\phi^L_{34}-\phi^L_{45}$.
In order to ensure the correctness of the above approximation, in Fig.(\ref{numvsann}) we plot the normalized absorption of the probe field (Im($\rho_{12}$)$\Gamma_2/\Omega_{12}$) as a function of normalized probe detuning ($\delta_{12}/\Gamma_2$) obtain
from analytical expression of $\rho_{12}$ as given in Eq.(\ref{ana}) as well as the complete numerical solution of density
matrix equation as stated in Eq.(\ref{ME}).
It is clear from Fig.(\ref{numvsann}) that the complete numerical results are in an excellent agreement with the approximated analytical solutions.
 
\section{RESULTS AND DISCUSSIONS}
\label{RESULTS}
\subsection{Lineshape of the probe absorption}
\label{RESULTS1}

\begin{figure}
\begin{center}
\includegraphics[width =\linewidth]{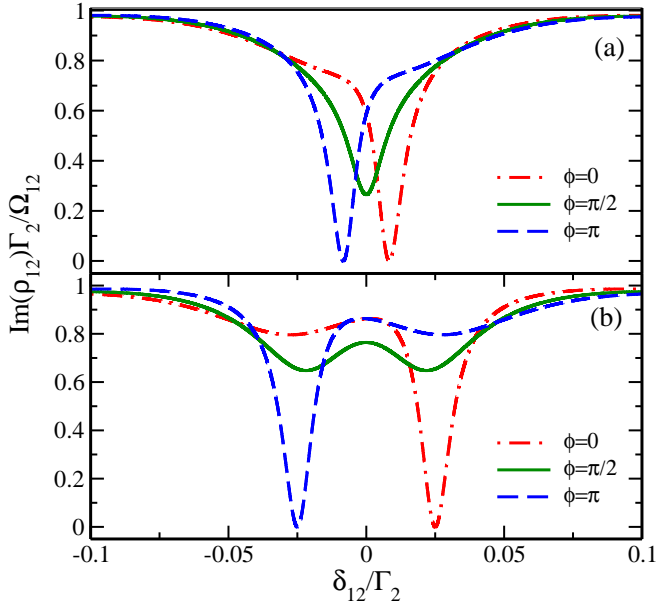}
\caption{(Color online). Normalized probe absorption Im($\rho_{12}$)$\Gamma_2/\Omega_{12}$ vs probe detuning in unit of $\Gamma_2$ ($\delta_{12}/\Gamma_2$) for (a) $|\Omega^{MW}_{35}|$=$2\pi\times$0.1~MHz (b) $|\Omega^{MW}_{35}|$=$2\pi\times$0.3~MHz. Other parameters are $\Omega_{12}$=$2\pi\times$0.01~MHz, $|\Omega_{23}|$=$|\Omega_{34}|$=$|\Omega_{45}|$=$2\pi\times$1~MHz $\delta_{23}$=$\delta_{34}$=$\delta_{45}$=$\delta^{MW}_{35}$=0. The dashed dotted red, solid green and dashed blue curves represent $\phi=0$, $\phi=\pi/2$ and $\phi=\pi$ respectively}
\label{probedet}
\end{center}
\end{figure}

In this section, we study the effect of the MW field on the lineshape of the probe absorption. 
We consider all the control and MW fields are on the resonance. 
First we discuss the role of individual control fields one by one and finally the MW field. 
There is a broad Lorentzian profile having a full width at half maximum of $\Gamma_2$ which is $2\pi\times$ 6~MHz in this case
\cite{IKN08, IFN09}. 
At the line centre of the Lorentzian profile, the probe response develops a sharp EIT dip in the presence of the control
field $\Omega_{23}$ \cite{IKN08,PAN13}.
The linewidth of EIT dip depends decoherence rates and the control field intensity
$|\Omega_{23}|$ and can be expressed as $\Delta\omega\propto|\Omega_{23}|^2/4 \Gamma_2$ \cite{Line}. 
Here we have taken $\Gamma_{1}$=$\Gamma_{3}$=0, as states $\ket{1}$ and $\ket{3}$ are the ground states.
The control laser $\Omega_{34}$ recovers the absorption against EIT and known as
EITA has been discussed in details in our previous work \cite{PAN13}.
When $|\Omega_{34}|<|\Omega_{23}|$ the linewidth of EITA peak is comparable with the linewidth of EIT dip and it is
broadening with the increase of the $\Omega_{34}$. Finally when $|\Omega_{34}|>|\Omega_{23}|$ the linewidth of the EITA peak
will be $\Gamma_2$.  
In the presence of the control laser $\Omega_{45}$ there will be transparency against EITA, known as EITAT and has also been
discussed in our previous work \cite{PAN13}.
The linewidth of the EITAT is given by the expression,

$\Delta\omega\propto\frac{|\Omega_{23}||\Omega_{45}|}{4\Gamma_2\left[(|\Omega_{34}|^2+|\Omega_{45}|^2)\left(\frac{1}{|\Omega_{23}|^2}+\frac{1}{|\Omega_{45}|^2}+\frac{|\Omega_{34}|^2}{|\Omega_{23}|^2|\Omega_{45}|^2}\right)-1\right]}$

The linewidth of the EITAT is modulated
by the three control field intensities $\Omega_{23}$, $\Omega_{34}$, $\Omega_{45}$ and the decay rate of the excited states
$\ket 2$, $\ket 4$. Here we have taken $\Gamma_{5}$=0, as state $\ket{5}$ is the metastable ground state.
The MW field, $\Omega^{MW}_{35}$ splits or shifts the EITAT dip, depending upon the phase $\phi$ as below as shown
in Fig. \ref{probedet}.
      
Fig. \ref{probedet} shows the plot for normalized absorption (Im($\rho_{12})\Gamma_2/\Omega_{12}$) vs probe detuning in the unit
of ($\Gamma_2$) for various combination of the phase and the MW field strength. In Fig. \ref{probedet}a for $\phi$=0 there is
shift of EITAT dip by approximately $\frac{|\Omega^{MW}_{35}|}{2}$ to the
right while it shifts to the left by same amount for $\phi$=$\pi$. For the $\phi=\pi/2$ there is broadening and
reduction of the EITAT dip. However with the relatively high value of the MW field, $|\Omega^{MW}_{35}|$=$2\pi\times$0.3~MHz
there is further reduction of EITAT dip and splitting increases and visible as shown in Fig. \ref{probedet}b.

\subsection{Effect of the MW power}
\label{RESULTS2}
\begin{figure}
\begin{center}
\includegraphics[width=\linewidth]{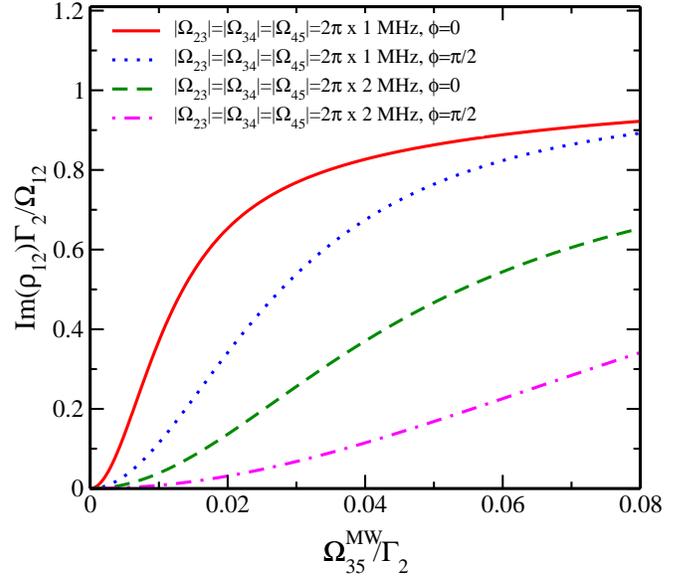}
\caption{(Color online). Im($\rho_{12}$)$\Gamma_2$/$\Omega_{12}$ vs $\Omega^{MW}_{35}$/$\Gamma_2$ for $\Omega_{12}$=$2\pi\times$0.01~MHz. The solid red curve is with $|\Omega_{23}|
=|\Omega_{34}|=|\Omega_{45}|$=$2\pi\times$1~MHz and $\phi=0$. The dashed green curve is with  $|\Omega_{23}|=|\Omega_{34}|$= $|\Omega_{45}|$=$2\pi\times$2~MHz and $\phi=0$.
The dotted blue curve is with $|\Omega_{23}|=|\Omega_{34}|$=$|\Omega_{45}|$=$2\pi\times$1~MHz and $\phi=\pi/2$. The magenta dashed dotted curve is with $|\Omega_{23}|=
|\Omega_{34}|$=$|\Omega_{45}|$=$2\pi\times$2~MHz and $\phi=\pi/2$.}
\label{VaryRFpower}
\end{center}
\end{figure}

The variation of the normalized probe absorption as a function of the MW Rabi frequency, $|\Omega^{MW}_{35}|$ in unit of 
$\Gamma_2$ is shown in Fig. \ref{VaryRFpower}. As in the presence of the MW field the EITAT dip shift,
thus with increase of the MW Rabi frequency the absorption increases and saturates to 1. 
The slope of the absorption profile depends upon the EITAT linewidth, and the narrower will be the linewidth sharper will be the slope. As we see in Fig. \ref{VaryRFpower} the red solid curve shows the sharper rise  in the absorption with the MW
Rabi frequency as compared to the dashed green curve. This is because the EITAT dip linewidth is power broadened by the control laser with higher Rabi frequency. The red
solid curve saturates faster as compare to the dotted blue curve this is because with $\phi=\pi/2$ the EITAT dip splits instead of shifting and hence will have lower slope
as compare to the $\phi=0$ with same Rabi frequency of the control lasers.
     
\begin{figure}
\begin{center}
\includegraphics[width =\linewidth]{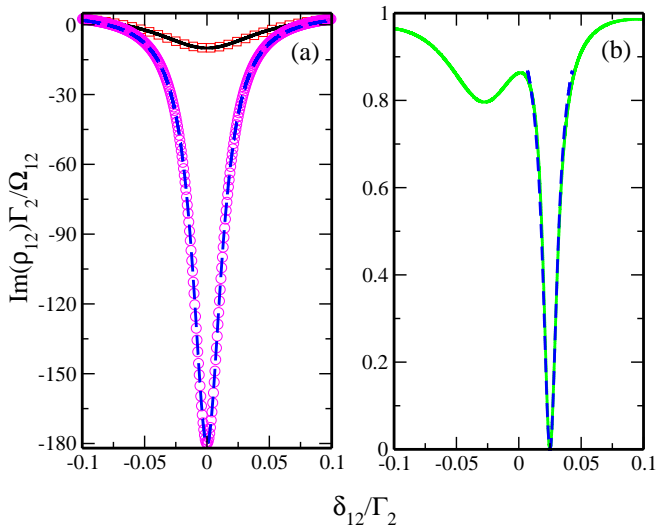}
\caption{(Color online). Normalized absorption, Im($\rho_{12}$)$\Gamma_2/\Omega_{12}$ as a function of probe detuning,
$\delta_{12}$ in unit of $\Gamma_2$ for (a) the closed loop $\Lambda$-system with $|\Omega_{12}|$=$2\pi\times$0.01~MHz,
$|\Omega_{23}|$=$2\pi\times$1~MHz, $|\Omega^{MW}_{35}|$=$2\pi\times$0.3~MHz and $\phi=\pi/2$ (b) the M-system
in which the parameters
are $|\Omega_{12}|$=$2\pi\times$0.01~MHz, $|\Omega_{23}|$=$|\Omega_{34}|$=$|\Omega_{45}|$=$2\pi\times$1~MHz,
$|\Omega^{MW}_{35}|$=$2\pi\times$0.3~MHz,
$\delta_{23}$=$\delta_{34}$=$\delta_{45}$=$\delta^{MW}_{35}$=0 and $\phi=0$.}
\label{probedet1}
\end{center}
\end{figure}

\section{closed loop $\Lambda$- vs M-system}
In this section, we discuss how the constraints of a closed loop $\Lambda$-system \cite{LVY09} can be overcome by
considering a M-system in which the MW field couples the unpopulated ground states. In Fig. \ref{probedet1}, we study
the probe absorption lineshape as a function of detuning under three-photon resonance condition i.e.
$\delta_{12}-\delta_{23}-\delta_{13}=0$ for a closed loop $\Lambda$-system and a M-system.
The lineshape under the approximation of no population transfer \cite{LVY09} is denoted by open circles in
Fig. \ref{probedet1}(a). A Lorentzian fit of the same (dashed blue curve) gives a linewidth of 0.028~$\Gamma_2$.
A full numerical solution for this system displays a broader linewidth (0.1~$\Gamma_2$) as shown by the open sqaures.
The broadening of the lineshape is due to the population transfer to the excited states which leads to fluorescence.
The solid green curve in Fig. \ref{probedet1}(b) shows the full numerical solution of the density matrix equation (4)
at steady state condition for M-system. It is found that the Lorentzian linewidth of the absorption spectrum is
0.014 $\Gamma_2$ as shown by dashed blue curve. Therefore the Lorentzian linewidth of M-system is much smaller than the
Lorentzian linewidth of closed loop $\Lambda$-system. Hence the population transfer can be suppressed in the former system
by MW field which is not possible in the latter. Nevertheless, the transparency window can be shifted by changing the total phase of the system at a fixed linewidth. Thus M-systems driven by a MW field can efficiently select the carrier frequency of the probe field.
\section{Conclusions}
\label{Conclusions}
In conclusion, we have studied how the strength and phase of the MW field can be efficiently used to manipulate the absorption property of the probe field in a five-level M-type configuration.
As our first result, we have found that the transparency window of the probe field can be shifted to a higher value by an amount $\frac{|\Omega^{MW}_{35}|}{2}$ under zero total phase $\phi$ and resonant control field. 
The  changes in the position of the transparency window in frequency space can enable us to select the desired carrier probe field frequency.
We also found that the linewidth of the probe spectrum can be changed by changing the strength and phase of both the control and MW fields.
Finally we have shown how M-systems can offer narrower linewidth of the probe absorption spectrum even at  strengths of MW field as high as the control field. This is unachievable in closed loop $\Lambda$-systems.
The absence of MW field induced population transfer in the M-systems can lead to this narrowing effect.
Hence this narrowing of probe absorption spectrum can be used for an atom based phase sensitive MW magnetometry.

\section*{Acknowledgment}
N.S.M would like to acknowledge financial support from MHRD, government of India . 

\bibliography{eitrefsall}

\end{document}